\documentclass[11pt]{article}
\usepackage{graphicx}
\usepackage{float} 
\usepackage{amsmath}
\usepackage{amsfonts}   
\usepackage{amssymb}

\begin{document}
\date{}
\title{{\bf{\Large Thermodynamics of phase transition in higher dimensional AdS black holes}}}
\author{
{\bf {{\normalsize Rabin Banerjee}}$
$\thanks{e-mail: rabin@bose.res.in}}
,$~${\bf {{\normalsize Dibakar Roychowdhury}}$
$\thanks{e-mail: dibakar@bose.res.in}}\\
 {\normalsize S.~N.~Bose National Centre for Basic Sciences,}
\\{\normalsize JD Block, Sector III, Salt Lake, Kolkata-700098, India}
\\[0.3cm]}



\maketitle


\begin{abstract}
We investigate the thermodynamics of phase transition for $ (n+1) $ dimensional Reissner Nordstrom (RN)-AdS black holes using a grand canonical ensemble. This phase transition is characterized by a discontinuity in specific heat. The phase transition occurs from a lower mass black hole with negative specific heat to a higher mass black hole with positive specific heat. By exploring Ehrenfest's scheme we show that this is a second order phase transition. Explicit expressions for the critical temperature and critical mass are derived. In appropriate limits the results for $ (n+1) $ dimensional Schwarzschild AdS black holes are obtained. 
\end{abstract}

\section{Introduction}
Black holes are exotic objects in the theory of classical and quantum gravity. Even more surprising is their connection with the laws of standard thermodynamics \cite{bch}. Since black hole thermodynamics is expected to play a role in any meaningful theory of gravity, therefore it will be a natural question to ask whether the thermodynamic properties of black holes are modified if higher dimensional corrections are incorporated in the Einstein-Hilbert action. One can expect a similar situation to appear in an effective theory of quantum gravity, such as string theory.  
Thermodynamics of black holes in higher dimensional AdS space has been a fascinating topic of research since the discovery of a phase transition by Hawking and Page in $ (3+1) $ dimensional Schwarzschild AdS background \cite{hp1}. Since then, a large number of investigations have been focused on the question of thermodynamic properties in higher dimensional anti de Sitter space time \cite{hp2}-\cite{hp18}. However none of these attempts rely on the conventional (gibbsian) way of discussing phase transitions in usual thermodynamic systems.

The approach to phase transitions in standard thermodynamics is based on the Clausius-Clapeyron-Ehrenfest's equations \cite{stanley,zeman}. These equations play an important role in order to classify the nature (first order or continuous (higher order)) of the phase transitions in standard thermodynamic systems. In spite of its several applications in various systems \cite{zeman}, \cite{glass1}-\cite{Jack}, it has never been systematically addressed to the phase transition phenomena in black holes. An attempt along this direction has been commenced recently \cite{bss}-\cite{dibakar}, where different black holes in AdS space were considered. But the analysis was strictly confined to $ (3+1) $ dimensions only. 

In this paper we adopt a gibbsian approach in the grand canonical frame work to discuss the thermodynamics of phase transition in arbitrary $ (n+1) $ dimensional RN AdS and Schwarzschild AdS black holes. This phase transition is characterized by a discontinuity of specific heat, compressibility and volume expansivity at the critical point. Our analysis clearly reveals that such a transition takes place between a black hole with smaller mass (phase 1) to a black hole with larger mass (phase 2). The specific heat corresponding to phase 1 is negative, while it is positive for phase 2. Therefore this phase transition is essentially a transition between a thermodynamically unstable phase to a thermodynamically stable one. Explicit expressions for the critical temperature ($ T_0 $) and critical mass ($ M_0 $) (segregating large from small mass black holes) are determined.  We also study the nature of this phase transition using Ehrenfest's scheme of standard thermodynamics. Our analysis reveals the fact that such transitions are indeed second order equilibrium transitions. In order to do that we analytically check both the Ehrenfest's equations at the critical point. It is to be noted that in spite of infinite divergences of all the relevant thermodynamic quantities (e.g; specific heat, volume expansion coefficient and compressibility) near critical point, we have been able to carry out our analysis. 

It is important to point out the difference between the phase transition we study in this paper with the conventional Hawking-Page phase transition. Our analysis is strictly confined to the critical temperature ($ T_0 $) where the specific heat ($ C_{\Phi} $) acquires an infinite divergence. On the other hand Hawking-Page transition occurs at a temperature $ T_1 $ ($ T_1> T_0  $) where Gibbs free energy ($G$) changes its sign \cite{hp9}-\cite{hp10}. From our analysis we are able to fix both $ T_0 $ and $ T_1 $. For large dimensions ($ n\rightarrow \infty $) we show that $ T_0 $ and $ T_1 $ are identical.  

Let us briefly mention about the organization of our paper. In section 2 we discuss various qualitative as well as quantitative features of phase transition both for the RN AdS and Schwarzschild AdS black holes in $ (n+1) $ dimensions. In section 3 we analyze the nature of this phase transition using Ehrenfest's scheme. Finally we draw our conclusions in section 4.

\section{Phase transition in higher dimensional AdS black holes}

Reissner-Nordstr{\"o}m black holes are charecterised by their mass (M) and charge (Q). The solution for $ (n+1) $ dimensional RN-AdS space time with a negative cosmological constant $ (\Lambda=-n(n-1)/2l^2) $ is defined by the line element \cite{hp11},
\begin{equation}
ds^2 = -\chi dt^2+\chi ^{-1}dr^2+r^2 d\Omega^{2}_{n-1}
\end{equation} 
where,
\begin{equation}
\chi(r) = 1-\frac{m}{r^{n-2}}+\frac{q^2}{r^{2n-4}}+\frac{r^2}{l^2}.
\end{equation}
Here $ m $ is related to the ADM mass ($ M $ ) of the of the black hole as, $ (G=1) $
\begin{equation}
M=\frac{(n-1)\omega_{n-1}}{16 \pi}m ~~ ;~~ ~~ \omega_{n-1} =\frac{2\pi^{n/2}}{\Gamma(n/2)} 
\label{m1}
\end{equation}
where $ \omega_{n-1} $ is the volume of unit $ (n-1) $ sphere.
The parameter $ q $ is related to the electric charge $ Q $ as
\begin{equation}
Q=\sqrt{2(n-1)(n-2)}\left( \frac{q}{8 \pi}\right).
\label{q1}
\end{equation}
The entropy of the system is defined as
\begin{equation}
S=\frac{\omega_{n-1} r_{+}^{n-1}}{4}
\label{s1}
\end{equation}
where $ r_{+} $ is the radius of the outer event horizon defined by the condition $ \chi(r_+ )=0$.
The electrostatic potential difference between horizon and infinity is given by
\begin{equation}
\Phi=\frac{\sqrt{n-1}}{\sqrt{2(n-2)}}\frac{q}{r_{+}^{n-2}}.
\label{phi}
\end{equation} 
Using (\ref{q1}) and (\ref{s1}) we can further express (\ref{phi}) as
\begin{equation}
\Phi=\frac{4 \pi Q}{(n-2)\omega^{\frac{1}{n-1}}_{n-1}(4S)^{\frac{n-2}{n-1}}}
\label{phi1}
\end{equation}
From the condition $ \chi(r_+ )=0$ and using (\ref{m1}), (\ref{q1}), (\ref{s1}) and (\ref{phi1}) we can express the black hole mass as
\begin{equation}
M=\frac{(4S)^{\frac{n-2}{n-1}}\omega^{\frac{1}{n-1}}_{n-1}(n-1)}{16 \pi}\left[ 1+\frac{2n-4}{n-1}\Phi^{2}+\left(\frac{4S}{\omega_{n-1}}\right) ^{\frac{2}{n-1}}\right].
\label{M} 
\end{equation}
Using the first law of black hole mechanics $ dM=TdS+\Phi dQ $, the Hawking temperature may be obtained as
\begin{eqnarray}
T&=&\left(\frac{\partial M}{\partial S}\right) _{Q}\nonumber\\
&=& \frac{4^{\frac{n-2}{n-1}}\omega^{\frac{1}{n-1}}_{n-1}(n-2)}{16 \pi S^{\frac{1}{n-1}}}\left[ 1-\frac{2n-4}{n-1}\Phi^{2}\right] + \frac{n 4^{\frac{n}{n-1}}S^{\frac{1}{n-1}}}{16 \pi \omega^{\frac{1}{n-1}}_{n-1} }.
\label{T} 
\end{eqnarray}
At this stage, it is customary to mention that in the following analysis the various black hole parameters $M,~Q,~J,~S,~T$ have been interpreted as $\frac{M}{l^{n-2}},~\frac{Q}{l^{n-2}},~\frac{S}{l^{n-1}},~Tl$ respectively, where the symbols have their standard meanings. Also, with this convention, the parameter $ l $  does not appear in any of the following equations.
\begin{figure}[h]
\centering
\includegraphics[angle=0,width=6cm,keepaspectratio]{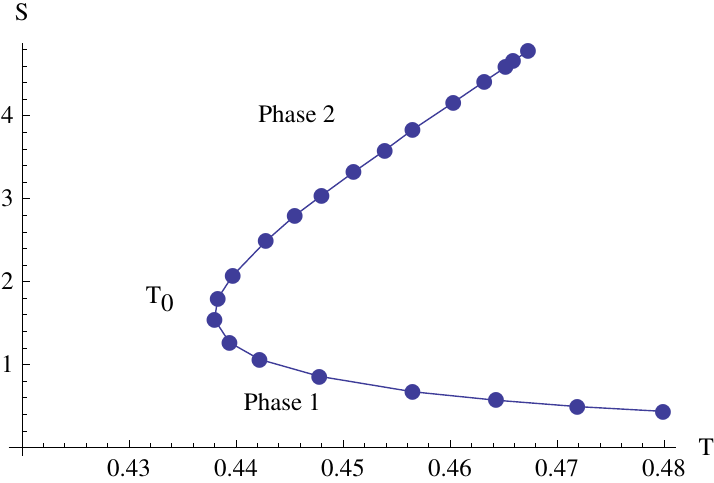}
\caption[]{\it Entropy plot ($ S $) for RN-AdS black hole with respect to temperature ($T$) for fixed $\Phi=0.2$ and $ n=4 $}
\label{figure 2a}
\end{figure} 

Using the above relation (\ref{T}) we plot the variation of entropy ($ S $) against temperature ($ T $) in figure 1. In order to obtain an explicit expression for the temperature ($ T_0 $) corresponding to the turning point we first set $ \left( \frac{\partial T}{\partial S}\right) _{\Phi} =0$. From this condition and using (\ref{T}) we find
\begin{equation}
S_{0}=\left(\frac{\omega_{n-1}}{4} \right)\left(\frac{n-2}{n} \right)^{(n-1)/2}\left[1-\frac{2n-4}{n-1}\Phi^{2} \right]^{(n-1)/2}.
\label{s0}
\end{equation}
Furthermore we find that 
\begin{equation}
\left[  \left( \frac{\partial^{2} T}{\partial S^{2}}\right) _{\Phi}\right] _{S=S_0}=\frac{n(n-2)}{8 \pi (n-1)^{3}}\left( \frac{4^{n}}{\omega_{n-1}S^{2n-3}}\right)^{1/(n-1)}.
\end{equation}
Therefore $\left( \frac{\partial^{2} T}{\partial S^{2}}\right) _{\Phi}> 0 $ for $ n>2 $. Hence the temperature $ T_0 $ corresponding to the entropy $ S=S_0 $ represents the minimum temperature for the system .
Substituting (\ref{s0}) into (\ref{T}) we find the corresponding minimum temperature for RN-AdS space time to be
\begin{equation}
T_{0}=\frac{\sqrt{n(n-2)}}{2\pi}\left[1-\frac{2n-4}{n-1}\Phi^{2} \right]^{1/2}.
\label{T0} 
\end{equation} 
In the charge-less limit ($\Phi=0$), we obtain the corresponding (minimum) temperature
\begin{equation}
 T_0= \frac{\sqrt{n(n-2)}}{2\pi}
 \label{t0}
\end{equation}
for Schwarzschild AdS space time in $ (n+1) $ dimension. For $ n=3 $ equation (\ref{t0}) yields
\begin{equation}
T_0=\frac{\sqrt{3}}{2 \pi}
\end{equation}
which was obtained earlier by Hawking and Page in their original analysis{\footnote{Note that $l$ does not appear since as explained before, it has been appropriately scaled out.}}  \cite{hp1}. 

It is now possible to interpret the two branches of the $ S-T $ curve, separated by the point $ (S_0,T_0) $, corresponding to two phases. Since entropy is proportional to the mass of the black hole, therefore phase 1 corresponds to a lower mass black hole and phase 2 corresponds to a higher mass black hole. We also observe that both of these phases exist above the (minimum) temperature $ T=T_0 $. To obtain an expression for the mass ($ M_0 $) of the black hole separating the two phases  we substitute (\ref{s0}) into (\ref{M}) which yields,
\begin{equation}
M_{0}=\frac{\omega_{n-1}(n-2)^{(n-2)/2}}{8 \pi n^{n/2}}\left[1-\frac{2n-4}{n-1}\Phi^{2} \right]^{(n-2)/2} 
\left[ (n-1)^{2}+ 2(n-2)\Phi^{2}\right]. 
\label{mc}
\end{equation}
Phase 1 corresponds to a black hole with mass $ M<M_0 $ while phase 2 corresponds to a black hole with mass $ M>M_0 $.

In the charge-less limit ($ \Phi=0 $) we obtain the corresponding critical mass for the Schwarzschild AdS black hole in    $ (n+1) $ dimension, 
\begin{equation}
M_0=\frac{\omega_{n-1}(n-2)^{(n-2)/2}(n-1)^{2}}{8 \pi n^{n/2}}.
\label{M0}
\end{equation} 
Substituting $ n=3 $ into (\ref{M0}) yields 
\begin{equation}
M_0=\frac{2}{3\sqrt{3}}
\end{equation}
which is the result obtained earlier in \cite{hp1}. 

A deeper look at the $ S-T $ plot further reveals that the slope of the curve gradually changes it's sign around $ T=T_0 $, thereby clearly indicating the discontinuity in the specific heat ($ C_{\Phi}=T(\partial S/ \partial T)_{\Phi} $) that is associated with the occurrence of a continuous higher order transition at $T=T_0 $. All these features become more prominent from the following analysis. In order to do that we first compute the specific heat at constant potential ($C_{\Phi}$), which is the analog of $C_{P}$ (specific heat at constant pressure) in conventional systems. This is found to be
\begin{eqnarray}
C_{\Phi}&=& T\left( \partial S/\partial T\right) _{\Phi}\nonumber\\ 
&=&\frac{(n-1)S\left[(n-2)\omega^{\frac{2}{n-1}}_{n-1}(1-\frac{2n-4}{n-1}\Phi^{2})+n(4S)^{\frac{2}{n-1}}\right] }{\left[ n(4S)^{\frac{2}{n-1}}-(n-2)\omega^{\frac{2}{n-1}}_{n-1}(1-\frac{2n-4}{n-1}\Phi^{2})\right] }. \label{cphi}
\end{eqnarray}

\begin{figure}[h]
\centering
\includegraphics[angle=0,width=7cm,keepaspectratio]{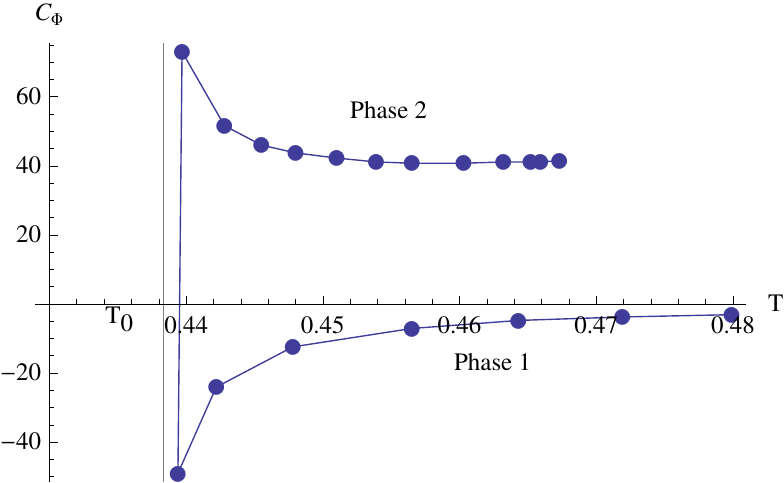}
\caption[]{\it Specific heat ($ C_{\Phi} $) plot for RN-AdS black hole with respect to temperature ($T$) for fixed $\Phi=0.2$ and $ n=4 $}
\label{figure 2a}
\end{figure} 

The critical temperature (at which the heat capacity ($ C_{\Phi} $) diverges) may be found by setting the denominator of (\ref{cphi}) equal to zero, which exactly yields the value of entropy ($ S $) as found in (\ref{s0}). Substituting this value into (\ref{T}) it is now trivial to show that $ T_0 $ (\ref{T0}) corresponds to the critical temperature at which $ C_{\Phi} $ diverges. This diverging nature of $C_{\Phi}$ at $ T=T_0 $ may also be observed from figure 2 where we plot heat capacity ($ C_{\Phi} $) against the temperature ($ T $). From this plot we find that the heat capacity ($ C_{\Phi} $) changes from negative infinity to positive infinity at  $ T=T_0 $ \cite{hp19}. One can further note that the smaller mass black hole (phase 1) corresponds to an unstable since it posses a negative specific heat ($C_{\Phi}<0$), whereas the larger mass black hole falls into a stable phase (phase 2) as it corresponds to a positive heat capacity ($C_{\Phi}>0$).
\begin{figure}[h]
\centering
\includegraphics[angle=0,width=7cm,keepaspectratio]{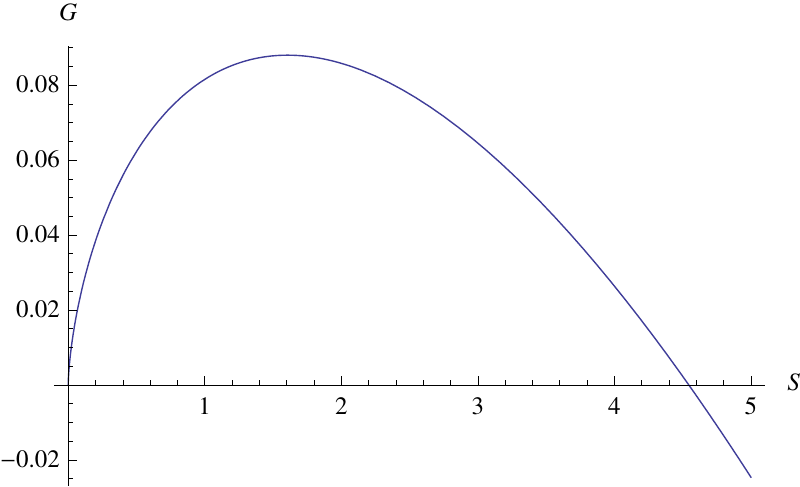}
\caption[]{\it Gibbs free energy ($G$) plot for RN-AdS black hole with respect to entropy ($S$) for fixed $\Phi=0.2$ and $ n=4 $}
\label{figure 2a}
\end{figure}

It is now customary to define grand canonical potential which is known as Gibbs free energy. For RN-AdS black hole this is defined as $G = M-TS-\Phi Q$ where the last term is the analog of $PV$ term in conventional systems. 
Using (\ref{phi1}),(\ref{M}),(\ref{T}) one can easily write $G$ as a function of ($S,~\Phi$),  
\begin{eqnarray}
G = \frac{(4S)^{\frac{n-2}{n-1}}\omega^{\frac{1}{n-1}}_{n-1}}{16 \pi}\left[ 1-\frac{2n-4}{n-1}\Phi^{2}\right] 
- \frac{ (4S)^{\frac{n}{n-1}}}{16 \pi \omega^{\frac{1}{n-1}}_{n-1} }.
\label{grn1}
\end{eqnarray}

Since in the above relation we cannot directly substitute entropy ($ S $) in terms of temperature ($ T $) therefore in order to obtain the $ G-T $ plot we first show the variation of Gibbs free energy ($ G $) against entropy ($ S $) using the above relation (\ref{grn1}).

As a next step, (from figure 3) we note the values of $ G $ for different values of $ S $ and obtain the corresponding values of $ T $ from figure 1. Finally we put all those data together in order to obtain the following numerical plot (fig. 4)

\begin{figure}[h]
\centering
\includegraphics[angle=0,width=7cm,keepaspectratio]{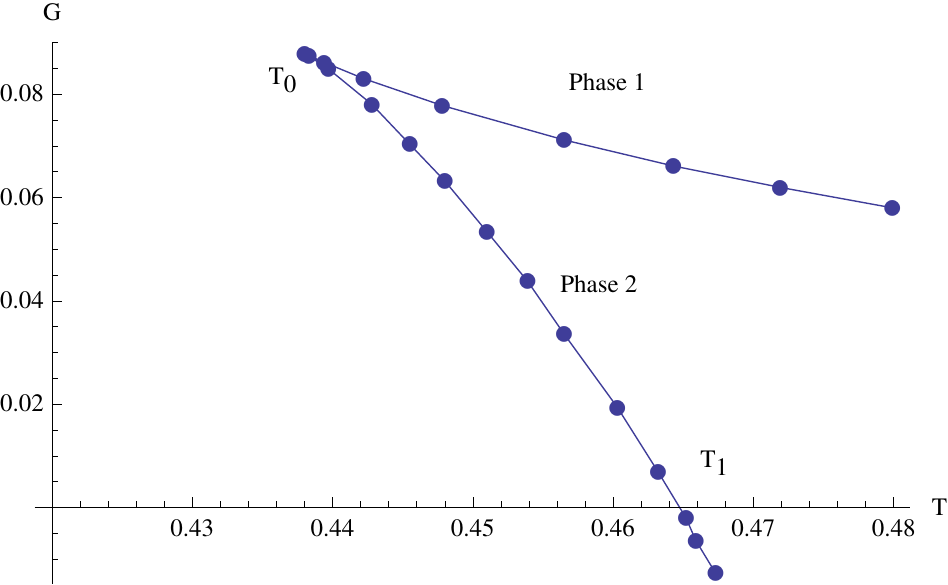}
\caption[]{\it Gibbs free energy ($G$) plot for RN-AdS black hole with respect to temperature ($T$) for fixed $\Phi=0.2$ and $ n=4 $}
\label{figure 2a}
\end{figure}

From figure 4 we note that the value of $ G $ for phase 1 is greater than that of phase 2. Therefore the larger mass black hole falls into a more stable phase than the black hole with smaller mass.  We also observe that $ G $ is always positive for $ T_0<T<T_1 $ . This also corroborates the findings of \cite{hp1} for the Schwarzschild AdS black hole.

In order to calculate $T_1$ we first set $ G=0 $ (in \ref{grn1}) which yields
\begin{equation}
S_1=\left( \frac{\omega_{n-1}}{4}\right) \left[ 1-\frac{2n-4}{n-1}\Phi^{2}\right]^ \frac{(n-1)}{2}.
\label{s'}
\end{equation}
Substituting this into (\ref{T}) we finally obtain
\begin{equation}
T_1=\frac{(n-1)}{2 \pi}\left[ 1-\frac{2n-4}{n-1}\Phi^{2}\right]^ \frac{1}{2}.
\label{Tg}
\end{equation}
In the charge-less limit ($ \Phi=0 $) we obtain the corresponding temperature for the Schwarzschild AdS black hole in    $ (n+1) $ dimension 
\begin{equation}
T_1=\frac{(n-1)}{2 \pi}.
\label{T'}
\end{equation}
Substituting $ n=3 $ into (\ref{T'}) yields 
\begin{equation}
T_1=\frac{1}{\pi}
\end{equation}
which reproduces the result found earlier in \cite{hp1}. 

Furthermore from (\ref{T0}) and (\ref{Tg}) we note that 
\begin{equation}
T_0=\frac{\sqrt{n(n-2)}}{n-1}T_{1}\label{univ}
\end{equation}
which is a universal relation in the sense that it does not depend on black hole parameters. Also note that for large value of $ n $ ($ n\rightarrow\infty $)
\begin{equation}
T_0=T_1=\frac{n}{2\pi}\left[1-2\Phi^{2}\right] ^{1/2}
\end{equation}
in the $ (n+1) $ dimensional RN-AdS back ground. Therefore this clearly suggests that the slope of the $ G-T $ plot (figure 4) becomes more steeper (for $ T_0<T<T_1 $) as we consider our theory in sufficiently higher dimensions. In the charge-less limit $ (\Phi=0 $) we obtain the corresponding result
 \begin{equation}
T_0=T_1=\frac{n}{2\pi}
\end{equation}
for the Schwarzschild AdS background in $(n+1)$ dimension.

\section{Phase transition and Ehrenfest's equations}
Classification of the nature of phase transition using Ehrenfest's scheme is a very elegant technique in standard thermodynamics. In spite of its several applications to other systems \cite{zeman}, \cite{glass1}- \cite{Jack}, it is still not a widely explored scheme in the context of black hole thermodynamics although some attempts have been made recently \cite{bss}-\cite{bgr}. In the remaining part of this paper we shall analyze and classify the phase transition phenomena in RN AdS black holes by exploiting Ehrenfest's scheme. 

Looking at the ($S-T$) graph (fig.1) we find that entropy $ (S) $ is indeed a continuous function of temperature $ (T) $. Therefore the possibility of a first order phase transition is ruled out.  However, the infinite divergence of specific heat near the critical point $ (T=T_0) $ (see fig.3) strongly indicates the onset of a higher order (continuous) phase transition. The nature of this divergence is further illuminated by looking at the $ C_{\Phi}-S $ graph.
\begin{figure}[h]
\centering
\includegraphics[angle=0,width=7cm,keepaspectratio]{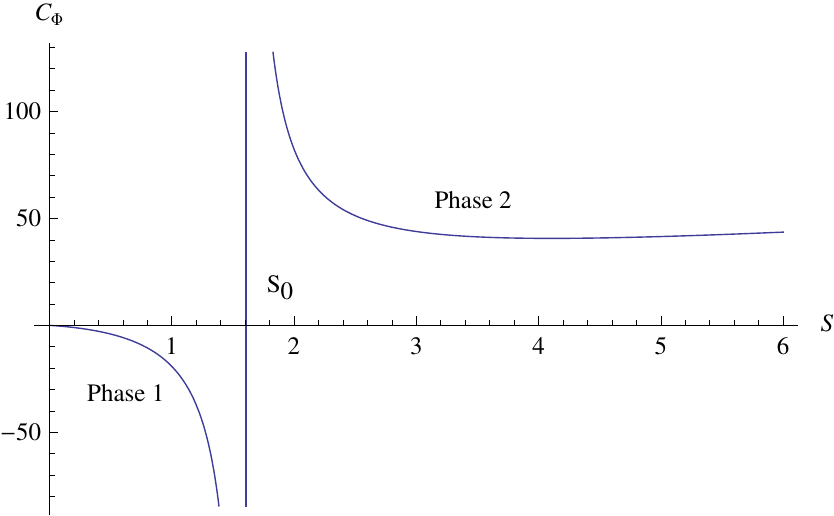}
\caption[]{\it Specific heat ($ C_{\Phi} $) plot for RN-AdS black hole with respect to entropy ($S$) for fixed $\Phi=0.2$ and $ n=4 $}
\label{figure 2a}
\end{figure}

We now exploit Ehrenfest's scheme in order to understand the nature of the phase transition.  Ehrenfest's scheme basically consists of a pair of equations known as Ehrenfest's equations of first and second kind \cite{stanley,zeman}. For a standard thermodynamic system these equations may be written as \cite{zeman}
\begin{eqnarray}
&& \left(\frac{\partial P}{\partial T}\right)_{S} = \frac{C_{P_2}-C_{P_1}}{T V(\alpha_2-\alpha_1)}=\frac{\Delta C_{P}}{T V \Delta \alpha}
\label{eh1}\\
&&\left(\frac{\partial P}{\partial T}\right)_{V} = \frac{\alpha_{2}-\alpha_{1}}{k_{T_{2}}-k_{T_{1}}}=\frac{\Delta \alpha}{\Delta k_T}
\label{eh2}
\end{eqnarray} 
For a genuine second order phase transition both of these equations have to be satisfied simultaneously.  

Let us now start by considering the RN-AdS black holes. Bringing the analogy ($ V \leftrightarrow Q, P \leftrightarrow -\Phi$) between the thermodynamic state variables and various black hole parameters, we are now in a position to write down the Ehrenfest's equations for this system as \cite{bgr} 
\begin{eqnarray}
&& -\left(\frac{\partial \Phi}{\partial T}\right)_{S} = \frac{C_{\Phi_2}-C_{\Phi_1}}{T Q(\alpha_2-\alpha_1)}=\frac{\Delta C_{\Phi}}{T Q \Delta \alpha}
\label{ehf1}\\
&&-\left(\frac{\partial \Phi}{\partial T}\right)_{Q} = \frac{\alpha_{2}-\alpha_{1}}{k_{T_{2}}-k_{T_{1}}}=\frac{\Delta \alpha}{\Delta k_T}
\label{ehf2}
\end{eqnarray} 
where, $\alpha =\frac{1}{Q}\left( \frac{\partial Q}{\partial T}\right) _{\Phi}$ is the analog of volume expansion coefficient and $k_T =\frac{1}{Q}\left( \frac{\partial Q}{\partial \Phi}\right) _{T}$ is the analog of isothermal compressibility. Their explicit forms are given by,\\ 
\begin{eqnarray}
\alpha &=& \frac{16 \pi (n-2)\omega^{\frac{1}{n-1}}_{n-1}S^{\frac{1}{n-1}}}{4^{\frac{n-2}{n-1}}\left[ n(4S)^{\frac{2}{n-1}}-(n-2)\omega^{\frac{2}{n-1}}_{n-1}(1-\frac{2n-4}{n-1}\Phi^{2})\right]}\label{alpha} \\
k_T &=& \frac{\left[ n(4S)^{\frac{2}{n-1}}-(n-2)\omega^{\frac{2}{n-1}}_{n-1}(1-\frac{(4n-6)(n-2)}{n-1}\Phi^{2})\right]}{\Phi \left[ n(4S)^{\frac{2}{n-1}}-(n-2)\omega^{\frac{2}{n-1}}_{n-1}(1-\frac{2n-4}{n-1}\Phi^{2})\right]}\label{kt}
\end{eqnarray}

Observe that the denominators of $ C_{\Phi} $ (\ref{cphi}), $ \alpha $ (\ref{alpha}), and $ k_T $ (\ref{kt}) are all identical. Hence $C_{\Phi}$, $ \alpha $ and $ k_T $ diverge at the critical point since, as already discussed, this point just corresponds to the vanishing of the denominator in $ C_{\Phi} $. However as shown below, the R.H.S. of (\ref{ehf1}) and (\ref{ehf2}) which involves the ratio of these quantities, remain finite at the critical point.

Using (\ref{T}), the L.H.S. of the first Ehrenfest's equation (\ref{ehf1}) may be found as (at the critical point $S_0$)  
\begin{equation}
-\left[ \left(\frac{\partial \Phi}{\partial T}\right)_{S}\right] _{S=S_0}= \frac{\pi (n-1)}{\Phi(n-2)^2}\left( \frac{4S_0}{\omega_{n-1}}\right) ^{1/(n-1)}.
\label{eh1}
\end{equation} 
In order to calculate the right hand side, we first note that
\begin{equation}
Q\alpha=(\partial Q/\partial T)_{\Phi}=(\partial Q/ \partial S)_{\Phi} (C_{\Phi}/T)
\end{equation}
Therefore the R.H.S. of (\ref{ehf1}) becomes
\begin{equation}
\frac{\Delta C_{\Phi}}{T_0 Q \Delta \alpha}=\left[ \left( \frac{\partial S}{\partial Q}\right) _{\Phi}\right] _{S=S_0}.
\label{Eq2}
\end{equation} 
Using (\ref{phi1}) we calculate the R.H.S. of the above equation and obtain,
\begin{equation}
\frac{\Delta C_{\Phi}}{T_0 Q \Delta \alpha}
=\frac{\pi (n-1)}{\Phi(n-2)^2}\left( \frac{4S_0}{\omega_{n-1}}\right) ^{1/(n-1)}.
\label{Eh2}
\end{equation}
Equations (\ref{eh1}) and (\ref{Eh2}) clearly reveal the validity of the first Ehrenfest's equation. Remarkably we find that the divergence in $C_{\Phi}$ is canceled with that of $\alpha$ in the first equation.

In order to evaluate the L.H.S. of (\ref{ehf2}) we first note that, since $ T=T(S,\Phi) $ , therefore,
\begin{equation}
\left(\frac{ \partial T}{\partial \Phi}\right) _{Q}=\left(\frac{ \partial T}{\partial S}\right) _{\Phi}\left(\frac{ \partial S}{\partial \Phi}\right) _{Q}+\left(\frac{ \partial T}{\partial \Phi}\right) _{S}.
\label{Eq1}
\end{equation} 
Since specific heat $ (C_{\Phi}) $ diverges at the critical point ($ S_{0} $) therefore it is evident from (\ref{cphi})that $\left[\left(\frac{\partial T}{\partial S}\right)_{\Phi}\right]_{S = S_{0}} = 0$. Also from (\ref{phi1}) we find that, $ \left(\frac{ \partial S}{\partial \Phi}\right) _{Q} $ has a finite value when evaluated at $ S=S_{0} $. Therefore the first term on the R.H.S. of (\ref{Eq1}) vanishes. This is a very special thermodynamic feature of AdS black holes, which may not be true for other systems. Therefore under this circumstances we obtain,
 \begin{equation}
 -\left[ \left(\frac{\partial \Phi}{\partial T}\right)_{Q}\right] _{S=S_0}=-\left[ \left(\frac{\partial \Phi}{\partial T}\right)_{S}\right] _{S=S_0}
 \label{Eq3}
 \end{equation}
Using the thermodynamic identity
\begin{equation}
\left(\frac{\partial Q}{\partial \Phi}\right)_{T}\left(\frac{\partial \Phi}{\partial T}\right)_{Q}\left(\frac{\partial T}{\partial Q}\right)_{\Phi}=-1
\end{equation}
we find, 
\begin{equation}
Qk_T=\left(\frac{\partial Q}{\partial \Phi}\right)_{T}=-\left(\frac{\partial T}{\partial \Phi}\right)_{Q}\left(\frac{\partial Q}{\partial T}\right)_{\Phi}=-\left(\frac{\partial T}{\partial \Phi}\right)_{Q} Q\alpha.
\end{equation}
Therefore R.H.S. of (\ref{ehf2}) may be found as
\begin{equation}
\frac{\Delta \alpha}{\Delta k_T}=-\left[ \left(\frac{\partial \Phi}{\partial T}\right)_{Q}\right] _{S=S_0}.
\label{Eq4}
\end{equation}
This shows the validity of the second Ehrenfest's equation. Also we find that divergences in $\alpha$ and $k_T$  get canceled in the second equation like in the previous case. 
Using (\ref{ehf1},\ref{Eq3},\ref{Eq4}) the Prigogine- Defay (PD) ratio $(\Pi) $ may be found to be
\begin{equation}
\Pi=\frac{\Delta C_{\Phi} \Delta k_T}{T_0 Q (\Delta\alpha)^{2}}=1.
\end{equation} 
Hence the phase transition occurring at $ T=T_0 $  is a second order equilibrium transition \cite{glass1,glass2},\cite{glass4,Jack}. This is true in spite of the fact that the phase transition curves are smeared and divergent near the critical point.

\section{Conclusions}
In this paper, based on standard thermodynamics, we have systematically developed an approach to analyze the phase transition phenomena in Reissner Nordstrom AdS black holes in arbitrary dimensions. In suitable limits the results for the Schwarzschild AdS black hole are obtained. This phase transition is characterized by divergences in specific heat, volume expansivity and compressibility near the critical point.  Furthermore, within the gibbsian approach, and using a grand canonical ensemble, we have provided a unique way (which is valid in any dimension greater than two) to analyze both the quantitative and qualitative features of this phase transition. From our analysis it is clear that such a transition takes place from a black hole with smaller mass ($ M<M_0 $) to a black hole with larger mass ($ M>M_0 $), where $ M_0 $ denotes the critical mass. Also, we have found that the black hole with smaller mass falls in an unstable phase since it has negative specific heat, whereas that with the larger mass possesses a positive specific heat and thereby corresponds to a stable phase. We have provided completely generalized expressions for the critical temperature ($ T_0 $) and critical mass ($ M_0 $) that characterize this phase transition. Also, we have explicitly calculated the temperature $ T_1 $ ($ T_1>T_0 $) which is associated with the change in sign in Gibb's free energy ($ G $). Following our approach, we have established a universal relation (\ref{univ}) between $ T_0 $ and $ T_1 $ that is valid in any arbitrary dimension ($ n>2 $).  Furthermore from our analysis, we have found that for large dimensions ($ n\rightarrow\infty $) both $ T_0 $ and $ T_1 $ become identical. 

Employing Ehrenfest's scheme we have resolved the vexing issue regarding the nature of phase transition in Reissner Nordstrom AdS black holes. In order to do that we have analytically checked the validity of both the Ehrenfest's equations near the critical point. This clearly suggests that the phase transition that is associated with the divergence in the heat capacity at the critical point ($ T_0 $) is indeed a second order equilibrium transition. 

Although there have been a number of investigations \cite{hp9}-\cite{hp11} in order to analyze the phase transition phenomena in higher dimensional (charged) AdS black holes, most of these are confined to the canonical ensemble. A detailed study based on a grand canonical frame work was lacking. Moreover, explicit expressions for the critical temperature ($ T_0 $) and/or critical mass ($ M_0 $), were in general, unavailable.

Our analysis reveals an important distinction between the thermodynamics of black holes and that of usual systems. While most discussions in the canonical ensemble  \cite{hp9}-\cite{hp10} point to a first order phase transition we obtain a second order transition in the grand canonical frame work. Thus, contrary to usual thermodynamics, the nature of the phase transition appears to depend on the specific ensemble. Our demonstration of the second order phase transition was based on proving the validity of the two Ehrenfest's relations near the critical point.  

{\bf{ Acknowledgement:}}\\
 D.R would like to thank the Council of Scientific and Industrial Research (C. S. I. R), Government of India, for financial help.



\end{document}